\pdfoutput=1
\RequirePackage{lineno}
\documentclass[
    aps,
    prd,
    twocolumn,
    superscriptaddress,
    amsmath,
    amssymb,
    nofootinbib,
    ]{revtex4-2}
\usepackage{graphicx} 
\usepackage{epstopdf} 
\usepackage{dcolumn} 
\usepackage{xcolor} 
\usepackage{amsmath,amssymb} 
\usepackage{hyperref} 
\usepackage[utf8]{inputenc} 
\usepackage[english]{babel} 
\usepackage{blindtext} 
\usepackage{ifthen} 
\usepackage{orcidlink} 

\usepackage{booktabs}       

\graphicspath{{figures/}} 

\newboolean{articletitles}
\setboolean{articletitles}{true} 

\input{belle2-symbols}

\newcommand{\BKnn}{\ensuremath{{B^{+}\to K^{+}\nu\bar{\nu}}}\xspace}

\def\BDT#1{\ensuremath{\mathrm{BDT}_#1}\xspace}

\def\combinationBFdetailed{\ensuremath{\left(2.3 \pm 0.5(\mathrm{stat})^{+0.5}_{-0.4}(\mathrm{syst})\right)\cdot 10^{-5}}\xspace}

\def\combinationsigSM{\ensuremath{2.7}\xspace}

\def\combinationsigO{\ensuremath{3.5}\xspace}

\newboolean{wordcount}
\setboolean{wordcount}{false} 

\ifthenelse{\boolean{wordcount}}%
{\usepackage{bibentry} 
 \usepackage{comment} 
 \excludecomment{align*} 
 \excludecomment{align} 
 \excludecomment{equation*} 
 \excludecomment{equation} 
 \excludecomment{eqnarray*} 
 \excludecomment{eqnarray} 
 \excludecomment{acknowledgments} 
 \def\maketitle{} 
}{}

\usepackage{xcolor}

\usepackage{makecell}
\usepackage[nameinlink, capitalise]{cleveref}

\begin{document}

\vspace*{-3\baselineskip}
\resizebox{!}{2cm}{\includegraphics{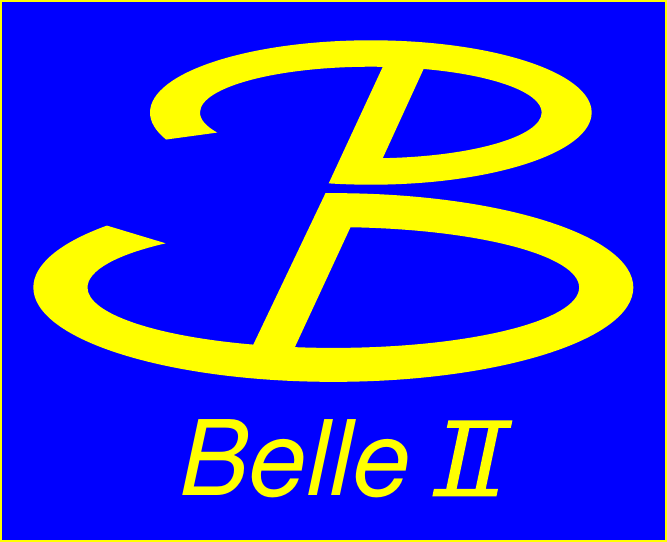}}


\begin{flushright}
\vspace*{24pt}
\onecolumngrid
\end{flushright}

\vspace*{24pt}

\title{
Model-agnostic likelihood for the reinterpretation \\ of the \texorpdfstring{\boldmath$B^+\to K^+\nu\bar{\nu}$}{B to K nu nu} measurement at Belle~II
}
  \author{M.~Abumusabh\,\orcidlink{0009-0004-1031-5425}} 
  \author{I.~Adachi\,\orcidlink{0000-0003-2287-0173}} 
  \author{L.~Aggarwal\,\orcidlink{0000-0002-0909-7537}} 
  \author{H.~Ahmed\,\orcidlink{0000-0003-3976-7498}} 
  \author{Y.~Ahn\,\orcidlink{0000-0001-6820-0576}} 
  \author{N.~Akopov\,\orcidlink{0000-0002-4425-2096}} 
  \author{S.~Alghamdi\,\orcidlink{0000-0001-7609-112X}} 
  \author{M.~Alhakami\,\orcidlink{0000-0002-2234-8628}} 
  \author{A.~Aloisio\,\orcidlink{0000-0002-3883-6693}} 
  \author{N.~Althubiti\,\orcidlink{0000-0003-1513-0409}} 
% ...
\collaboration{The Belle II Collaboration}

\begin{abstract}
We recently measured the branching fraction of the $B^{+}\rightarrow K^{+}\nu\bar{\nu}$ decay using 362~fb$^{-1}$ of on-resonance $e^+e^-$ collision data under the assumption of Standard Model kinematics, providing the first evidence for this decay. 
To facilitate future reinterpretations and maximize the scientific impact of this measurement, we publicly release the full analysis likelihood along with all necessary material required for reinterpretation under arbitrary theoretical models sensitive to this measurement.
In this work, we demonstrate how the measurement can be reinterpreted within the framework of the Weak Effective Theory. Using a kinematic reweighting technique in combination with the published likelihood, we derive marginal posterior distributions for the Wilson coefficients, construct credible intervals, and assess the goodness of fit to the Belle~II data.
For the Weak Effective Theory Wilson coefficients, the posterior mode of the magnitudes $|C_\mathrm{VL}+C_\mathrm{VR}|$, $|C_\mathrm{SL}+C_\mathrm{SR}|$, and $|C_\mathrm{TL}|$ corresponds to the point ${(11.3, 0.0, 8.2)}$. The respective 95\% credible intervals are $[1.9, 16.2]$, $[0.0, 15.4]$, and $[0.0, 11.2]$.

\end{abstract}

\maketitle

\section{Introduction}

Measurements are typically performed under the assumptions of a specific theoretical model, but have the potential to constrain a large variety of additional models.
Reinterpretations neglecting variations in detector efficiency, at scales below the provided resolution, can introduce biases. More robust alternatives typically require access to the full analysis strategy and the computationally intensive event simulation. Alternatively, the ability to perform event-level reweighting using the original analysis samples is needed. These are usually not publicly accessible.
The method introduced in Ref.~\cite{Gartner:2024muk} offers an efficient approach to reinterpret measurements in terms of alternative theoretical models, while accurately accounting for efficiency variations at high resolution. 
The key advantage of this method lies in its ability to construct and publish a reinterpretable, \textit{model-agnostic}, likelihood, which enables the assessment of data compatibility with a wide range of theoretical models. 
In this work, we provide the necessary information to apply this method to our $\BKnn$ result~\cite{Belle-II:2023esi} and demonstrate its potential using a selected alternative model.

Flavor-changing neutral-current $b\to s \nu\bar{\nu}$ transitions are suppressed in the Standard Model (SM) of particle physics due to the Glashow-Iliopoulos-Maiani mechanism \cite{Glashow:1970gm}. The branching fraction of the exclusive $b \to s \nu \bar{\nu}$ process, \BKnn decay,\footnote{Charge-conjugate channels are implied throughout this work.} is predicted in the SM to be~\cite{Parrott:2022zte}
\begin{equation}
\mathcal{B}(\BKnn) =
\left(5.58 \pm 0.37 \right)\cdot 10^{-6}\, ,
\label{eq:brf}
\end{equation}
which includes a contribution of $\left(0.61 \pm 0.06\right)\cdot 10^{-6}$ from the long-distance double-charged-current ${B^+\to \tau^+(\to K^+\bar{\nu}_\tau)\nu_\tau}$ decay. 

Since the $B$ meson is a pseudoscalar, the decay is isotropic in its rest frame. The only experimentally accessible kinematic degree of freedom is the squared dineutrino invariant mass, $q^2 =(p_\nu + p_{\bar \nu})^2 = (p_B - p_K)^2$, which is a key variable in probing new physics (NP) beyond the SM. The differential branching fraction $\text{d}\mathcal{B}/\text{d}q^2$ is directly proportional to the squared $B \to K$ transition form factor, $\left| f_+(q^2) \right|^2$, and the phase-space factor. The largest theoretical uncertainty in the SM prediction stems from imprecise knowledge of $f_{+}(q^2)$, which is parametrized by $3$ hadronic parameters.

The $\BKnn$ decay has been studied by the CLEO, \textit{BABAR}, Belle, and Belle~II collaborations~\cite{PhysRevLett.86.2950,PhysRevD.87.111103,PhysRevD.87.112005,PhysRevD.82.112002,PhysRevD.96.091101,PhysRevLett.127.181802,Belle-II:2023esi}, with the latest measurement by the Belle~II collaboration finding the first evidence for this decay at \combinationsigO\ standard deviations. 
This result, based on the SM prediction from Ref.~\cite{Parrott:2022zte} and hadronic parameters from the HPQCD collaboration~\cite{Parrott:2022rgu} as a model ansatz, exceeds the SM expectation by \combinationsigSM\ standard deviations.
This enhanced branching fraction triggered the high-energy physics community to interpret this result under different NP scenarios~\cite{PhysRevD.109.115006,PhysRevD.109.075008, Gabrielli:2024wys}. However, such reinterpretations are only approximate as not all relevant information on the measurement was previously accessible to the public.

Generally, two distinct classes of NP models can replicate the signature of a \BKnn decay. 
These models naturally mimic the neutrino pair in the final state of \BKnn decays, as the neutrinos remain undetected at Belle II. NP scenarios can proceed via either three-body or two-body decays.
In three-body decays, the properties of NP can be studied within the framework of the Weak Effective Theory (WET). This is an effective quantum field theory that describes both the SM and potential NP effects within a common parameter framework, where NP particles and force carriers have masses at or above the electroweak symmetry-breaking scale. Examples of such models include leptoquarks~\cite{PhysRevD.98.055003} and heavy $Z'$ bosons~\cite{PhysRevD.104.053007}.
Two-body decays would signal the presence of light NP, which lies below the scale of electroweak symmetry breaking, and is not described by the WET. Explicit examples include axion-like particles~\cite{PhysRevD.102.015023} or other dark-sector mediators~\cite{PhysRevD.101.095006}.

The primary goals of this work are twofold: to publish the model-agnostic likelihood for the \BKnn analysis; and to reinterpret the result in the framework of the WET, providing constraints on relevant Wilson coefficients.
This work includes: a summary of the most relevant aspects of the \BKnn analysis~\cite{Belle-II:2023esi}; a description of the kinematic reweighting method used for reinterpretation; details of the statistical inference procedure; results for the WET reinterpretation; and comparisons of model fit quality and relative performance with respect to the SM and background-only hypotheses.

\section{Overview of the Belle~II \texorpdfstring{$B^+\to K^+\nu\bar{\nu}$}{B to K nu nu} analysis}

The Belle~II analysis of \BKnn decays~\cite{Belle-II:2023esi}, originating from an $e^+ e^- \to \Upsilon(4S) \to B^+B^-$ process, was performed using two different reconstruction methods. These are the more sensitive \textit{inclusive} tagging analysis (ITA) and the more conventional \textit{hadronic} tagging analysis (HTA), targeting nearly orthogonal datasets. 
In the HTA, we reconstruct the non-signal $B$ meson of the $B^+B^-$ pair in specific hadronic decay modes. Further, the signal $B$ meson is reconstructed, providing a well-constrained event topology with low background but limited efficiency.
In contrast, in the ITA, we identified the signal by reconstructing the $K^{+}$ track, summing all visible particles in the event, and then inferring the missing energy. Thereby, we achieved higher sensitivity, though at the cost of increased background contamination. 
In addition to a basic preselection, the ITA used two boosted decision trees ($\mathrm{BDT}_1$ and $\mathrm{BDT}_2$) while the HTA used one ($\mathrm{BDTh}$) to separate the signal from the background. 
The statistical model was constructed using Monte Carlo simulated data for the signal and seven background categories: decays of charged and neutral $B \bar{B}$ mesons; continuum processes ($u \bar{u}$, $d \bar{d}$, $s \bar{s}$, $c \bar{c}$); and low-multiplicity $\tau^+ \tau^-$ backgrounds. In this measurement, only the short-distance contribution was considered as signal, with a corresponding branching fraction of $(4.97\pm 0.37) \cdot 10^{-6}$~\cite{Parrott:2022zte}.

For both the ITA and HTA methods, a likelihood was constructed using the \texttt{HistFactory} statistical model~\cite{histfactory}. 
In simplified form, this can be written as
\begin{equation}
    p\left( \boldsymbol{n}, \boldsymbol{a} | \boldsymbol{\eta}, \boldsymbol{\chi}\right) = p\left(\boldsymbol{n} | \boldsymbol{\nu}(\boldsymbol{\eta}, \boldsymbol{\chi})\right) ~ p(\boldsymbol{a} | \boldsymbol{\chi})\, ,
    \label{eq:histfactory}
\end{equation}
which is the product of a binned Poissonian likelihood and a constraint term.
Here, $\boldsymbol{n}$ and $\boldsymbol{a}$ are the observed and auxiliary data, respectively. The expected event counts $\boldsymbol{\nu}(\boldsymbol{\eta}, \boldsymbol{\chi})$ depend on the unconstrained parameters of interest $\boldsymbol{\eta}$ and the constrained nuisance parameters $\boldsymbol{\chi}$.

The analysis likelihood was implemented in the \texttt{pyhf} framework~\cite{pyhf, pyhf_joss}, incorporating all systematic uncertainties. The ITA analysis used $4 \times 3$ bins in the $\eta\left(\mathrm{BDT}_2\right) \times q^2_{\rm rec}$ space. Here, $\eta\left(\mathrm{BDT}_2\right)$ is the transformed $\mathrm{BDT}_2$ output, such that each bin represents a 2\% efficiency quantile for the selection. 

The variable $q^2_{\rm rec}$ is the reconstructed squared momentum transfer, defined as $q^2_{\rm rec} \equiv s/4 + M_K^2 - \sqrt{s} E_K^*$, where $\sqrt{s}$ is the center of mass energy, $M_K$ is the nominal kaon mass, and $E_K^*$ is the reconstructed energy of the kaon in the collision center of mass frame. The ITA and HTA $q^{2}$ resolutions are approximately $1.0\ \text{GeV}^{2}$ and $0.3\ \rm{GeV}^{2}$, respectively, leading to differences in the $q^{2}_{\rm rec}$ and $q^{2}$ distributions.\footnote{We set $c=1$ throughout this work.} Final state radiation has a negligible effect on the presented results.

The ITA includes both a signal and a control region, resulting in a total of 24 reconstruction bins. Signal region bins are populated with events that pass all selection criteria in 362~fb$^{-1}$ of on-resonance data. This data was taken at a center of mass energy of $\sqrt{s}\approx10.58$~GeV, which corresponds to the $\Upsilon(4S)$ resonance. Control region bins are filled with events that pass the same selection in 42~fb$^{-1}$ of off-resonance data, taken at $\sqrt{s} \approx 10.52$~GeV. The control region improves constraints on the background contributions from continuum processes.
The HTA used $6$ bins in the $\eta\left(\mathrm{BDTh}\right)$ space. These bins contain only events that pass all selection criteria in 362~fb$^{-1}$ of on-resonance data.

The ITA and HTA likelihoods were combined into one likelihood, accounting for the correlations between the systematic uncertainties in the two methods.
The \textit{combined} likelihood includes 231 nuisance parameters, in addition to one parameter of interest, the signal strength $\boldsymbol{\eta} = [\mu_{\rm SM}]$. This represents the signal branching fraction relative to its SM expectation. A maximum likelihood fit yielded a signal strength of ${\mu_{\rm SM} = 4.6 \pm 1.3}$, which corresponds to a branching fraction of \combinationBFdetailed. This result has a significance of 3.5 standard deviations over the background-only hypothesis and a significance of 2.7 standard deviations over the SM hypothesis. The combined likelihood serves as the basis for reinterpretation in this study.

\section{Reinterpretation method}

For the reinterpretation of such a SM-dependent template likelihood, an alternative signal template must be derived from the corresponding theoretical prediction, while ensuring consistency with the experimental acceptance. This is enabled by the reinterpretation method introduced in Ref.~\cite{Gartner:2024muk}, which allows the construction of a model-agnostic likelihood from the published \BKnn likelihood~\cite{Belle-II:2023esi} using histogram reweighting.
The key innovation of this method lies in its ability to capture the effect of reweighting a measured observable distribution based on a parametrized theoretical distribution. Crucially, this works without requiring access to the original event-level Monte Carlo simulation.

The basic idea is as follows. 
To interpret a measurement in terms of any theoretical model, we need the number density of expected events, $\nu(x)$, as a function of the observables $x$ used in the measurement.
This is obtained from the theoretical cross-section, $\sigma(q^2)$, which in this case depends on the kinematic degree of freedom  $q^2$,
\begin{equation}
    \nu(x) = L \int dq^2 ~ \varepsilon(x|q^2) ~ \sigma(q^2) = \int dq^2 ~ \nu(x,q^2)\, ,
\end{equation}
where $\varepsilon(x|q^2)$ is the combined reconstruction and selection efficiency, $L$ is the total integrated luminosity, and $\nu(x,q^2) \equiv L ~ \varepsilon(x|q^2) ~ \sigma(q^2)$ defines the joint number density.

To reinterpret the measurement in terms of a new theory, we need to determine the number density $\nu_1(x)$ for an \textit{alternative} theoretical prediction $\sigma_1(q^2)$. 
The original analysis uses the \textit{null} distribution, $\sigma_0(q^2)$, to obtain $\nu(x) \equiv \nu_0(x)$ from simulated events.
We obtain the alternative number density of expected events by reweighting,
\begin{equation}
    \nu_1(x) = \int dq^2 ~ \nu_0(x,q^2) ~ w(q^2)\, ,
    \label{eq:reweight}
\end{equation}
where $\nu_0(x,q^2)$ is the joint number density for the null distribution and $w(q^2) = \sigma_1(q^2) / \sigma_0(q^2)$ is the weight factor. 

The null distribution serves as our reference point for comparison. For the \BKnn analysis, this is the SM prediction from Ref.~\cite{Parrott:2022zte} with hadronic parameters from the HPQCD collaboration~\cite{Parrott:2022rgu}. The alternative distribution can be any beyond-SM or updated SM prediction.

For binned data (like the \BKnn likelihood~\cite{Belle-II:2023esi}), the reweighting step of \cref{eq:reweight} becomes a discrete sum:
\begin{equation}
    \nu_{1,x} = \sum_{q^2~\text{bins}} ~ \nu_{0,xq^2} ~ w_{q^2}\, ,
    \label{eq:reweight_discrete}
\end{equation} 
where the subscripts represent bin indices. The discrete quantities $\nu_{0,xq^2}$ and $w_{q^2} = \sigma_{1,q^2} / \sigma_{0,q^2}$ can be obtained from the continuous counterparts by integrating over bin intervals, as detailed in Ref.~\cite{Gartner:2024muk}.

Critically, this reweighting process only requires two pieces of information: the joint number density $\nu_{0,xq^2}$ from the original analysis, and the weight factor $w_{q^2}$ from the ratio of theoretical predictions.
Combined with the likelihood, these provide sufficient information to test alternative theories without access to the original Monte Carlo samples. This reinterpretation method, integrated within the \texttt{pyhf} framework \cite{pyhf, pyhf_joss}, is implemented in the \texttt{redist} software~\cite{redist_v1.0.4}.

The method has some limitations. It performs best when the alternative theory remains close to the null distribution. When theories deviate significantly, sparsely populated phase space regions receive very large weights, potentially leading to unreliable results. In particular, the alternative distribution should not extend beyond the kinematic range of the null distribution~\cite{Gartner:2024muk}.

\begin{figure}
    \centering
    \includegraphics[width=0.5\textwidth]{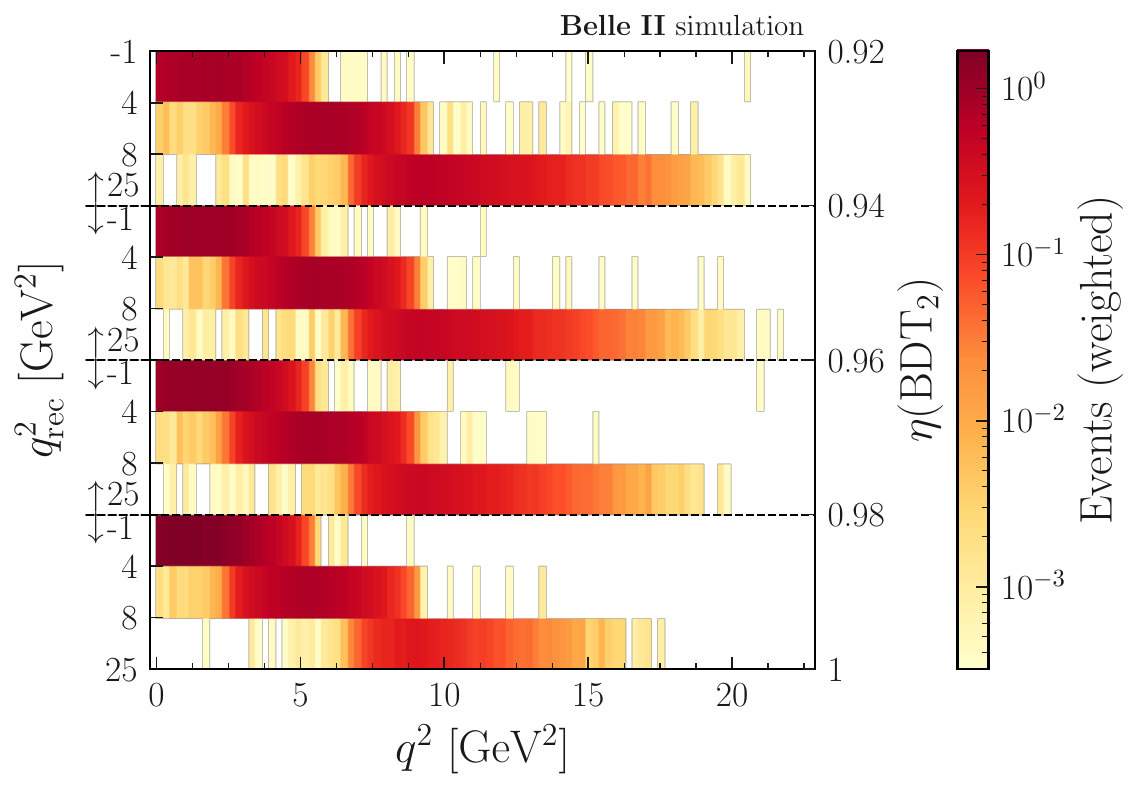}
    \caption{The ITA binned joint number densities. The horizontal axis corresponds to the generated $q^2$. The vertical axis represents the binning used in the \BKnn analysis~\cite{Belle-II:2023esi}.
    The heatmap shows the weighted signal events.}
    \label{fig:ita-map}
\end{figure}

To obtain the joint number density $\nu_{0,xq^2}$, we use simulated SM signal events from the \BKnn analysis~\cite{Belle-II:2023esi}, satisfying all selection criteria. These include information on the generated and reconstructed squared momenta, $q^2$ and $q^2_{\rm rec}$, as well as the classifier responses $\eta(\mathrm{BDT}_2)$ and $\eta(\mathrm{BDTh})$. The number of $q^2$ bins for $\nu_{0, x q^2}$ is determined by the differences between the null and the anticipated alternative distributions. 
The null distribution is the \BKnn SM prediction based on the form factors from Ref.~\cite{Parrott:2022rgu}.
The WET predicts a broad distribution in $q^2$. With future studies in mind, the binning strategy is optimized to capture localized features in the $q^2$ spectrum, resulting in 100 equally spaced $q^2$ bins in the kinematically allowed region plus one negative $q^2$ bin for events falling outside of this region. An example of $\nu_{0,xq^2}$ for the ITA is shown in \cref{fig:ita-map}.

In this reinterpretation study, we build on the \texttt{HistFactory} likelihood \cite{histfactory} to construct a Bayesian posterior for parameter inference. Starting from the likelihood in \cref{eq:histfactory}, we construct the posterior model
\begin{equation}
    \begin{aligned}
        p\left( \boldsymbol{\eta}, \boldsymbol{\chi} | \boldsymbol{n}, \boldsymbol{a} \right) \propto p\left(\boldsymbol{n} | \boldsymbol{\nu}(\boldsymbol{\eta}, \boldsymbol{\chi})\right) ~ p\left( \boldsymbol{\chi} | \boldsymbol{a} \right) ~ p\left( \boldsymbol{\eta} \right)\, .
    \end{aligned}
    \label{eq:BayesTheorem}
\end{equation}
The constraint likelihood is translated into a prior for all constrained parameters ${p(\boldsymbol{\chi}| \boldsymbol{a}) \propto p(\boldsymbol{a} | \boldsymbol{\chi}) ~ p(\boldsymbol{\chi})}$ with a normally distributed initial prior $p(\boldsymbol{\chi})$, as detailed in Ref.~\cite{Feickert:2023hhr}. 
Additionally, a prior for unconstrained parameters $p\left( \boldsymbol{\eta}\right)$ is introduced. 
The \texttt{bayesian pyhf} framework~\cite{Feickert:2023hhr} implements the posterior, using \texttt{pymc} \cite{AbrilPla2023PyMCAM} as a back-end for sampling.

\section{Weak Effective Theory}

A commonly-used framework to describe NP scenarios without using a UV-complete theoretical model is the WET.
The Lagrangian density for the $sb\nu\nu$ sector reads~\cite{Felkl:2021uxi,Gartner:2024muk}
\begin{equation}
    \mathcal{L}_\text{WET} \supset - \frac{4 G_\text{F}}{\sqrt{2}} \frac{\alpha}{2 \pi} V_{t s}^* V_{t b}
    \sum_i C_i(\mu_b) O_i + \text{h.c.}\,.
    \label{eq:wet-lagrangian}
\end{equation}
The operators $O_i$ describe interactions at energies below the separation scale $\mu_b = 4.2~{\rm GeV}$. The complex-valued Wilson coefficients $C_i(\mu_b)$ encode the dynamics above this energy scale, calculated in the $\overline{\text{MS}}$ scheme. In the SM, only $C_{\mathrm{VL}}$ is non-zero, with $C_{\mathrm{VL}}^{\rm SM}=6.6 \pm 0.1$~\cite{Parrott:2022zte}.
Here, $G_F$ is the Fermi constant, $\alpha$ is the fine structure constant, and $V_{ts}$ and $V_{tb}$ are elements of the Cabibbo-Kobayashi-Maskawa quark mixing matrix.

The full set of relevant dimension-six operators is~\cite{Felkl:2021uxi},
\begin{equation}
\begin{aligned}
\mathcal{O}_{\mathrm{VL}} &=\left(\overline{\nu_L} \gamma_\mu \nu_L\right)\left(\overline{s_L} \gamma^\mu b_L\right) \\
\mathcal{O}_{\mathrm{VR}} &=\left(\overline{\nu_L} \gamma_\mu \nu_L\right)\left(\overline{s_R} \gamma^\mu b_R\right) \\
\mathcal{O}_{\mathrm{SL}} &=\left(\overline{\nu_L^c} \nu_L\right)\left(\overline{s_R} b_L\right) \\
\mathcal{O}_{\mathrm{SR}} &=\left(\overline{\nu_L^c} \nu_L\right)\left(\overline{s_L} b_R\right) \\
\mathcal{O}_{\mathrm{TL}} &=\left(\overline{\nu_L^c} \sigma_{\mu \nu} \nu_L\right)\left(\overline{s_R} \sigma^{\mu \nu} b_L\right)\,.
\end{aligned}
\label{eq:operators}
\end{equation}
Here, $q_{L/R}$ and $\nu_{L/R}$ denote left- and right-handed quark and neutrino fields, respectively.
The charge-conjugated neutrino field is $\nu_L^c \equiv C \overline{\nu_L}^T$, where $C$ is the charge conjugation operator. The sigma tensor $\sigma^{\mu\nu} \equiv \frac{i}{2} [\gamma^\mu, \gamma^\nu]$ is the antisymmetric combination of Dirac gamma matrices. 
The subscripts V, S, T denote vector, scalar, and tensor operators, respectively. The analysis assumes massless neutrinos, with operators summed over all neutrino flavors.

The scalar and tensor operators are of dimension six in the WET and their matrix elements therefore enter at the same level as the vector operator. However, additional suppression is possible. For example, when matching to the Standard Model Effective Field Theory (SMEFT), the scalar and tensor WET coefficients receive contributions only from SMEFT operator of dimension seven or higher~\cite{Jenkins:2017jig}.

The resulting differential branching fraction for $\BKnn$ as predicted by the WET is given by~\cite{PhysRevD.93.054008,Felkl:2021uxi}
\begin{equation}
    \begin{aligned}
      \frac{d \mathcal{B}}{d q^{2}}
      & =
      3 \tau_B
      \left(\frac{4 G_\text{F}}{\sqrt{2}} \frac{\alpha}{2 \pi} \right)^2 \left|V_{t s}^* V^{}_{t b}\right|^2
      \frac{\sqrt{\lambda_{B K}} q^{2}}{(4 \pi)^{3} M_{B}^{3}}\\
      &\cdot\left[\frac{\lambda_{B K}}{24 q^{2}}\left|f_{+}(q^2)\right|^{2}\left|C_{\mathrm{VL}}+C_{\mathrm{VR}}\right|^{2}\right.\\
      &\phantom{\cdot}+\frac{\left(M_{B}^{2}-M_{K}^{2}\right)^{2}}{8\left(m_{b}-m_{s}\right)^{2}}\left|f_{0}(q^2)\right|^{2}\left|C_{\mathrm{SL}}+C_{\mathrm{SR}}\right|^{2} \\
      &\phantom{\cdot}\left.+\frac{2 \lambda_{B K}}{3\left(M_{B}+M_{K}\right)^{2}}\left|f_{T}(q^2)\right|^{2}\left|C_{\mathrm{TL}}\right|^{2}\right]\, ,
    \end{aligned}
\label{eq:wet-width}
\end{equation}
where $M_B$ and $M_K$ are the masses of the $B$ meson and the kaon, respectively. The quantities $m_b$ and $m_s$ are the masses of the $b$ and $s$ quarks in the $\overline{\text{MS}}$ scheme, respectively. The term ${\lambda_{B K} \equiv \lambda(M_B^2, M_K^2, q^2)}$ is the K\"all\'en function, and $\tau_B$ is the lifetime of the $B$ meson.

Due to sensitivity to only the absolute values of the three linear combinations of Wilson coefficients, this analysis treats each linear combination as a real-valued number.

The hadronic matrix elements are described by three independent hadronic form factors $f_{+}(q^2)$, $f_{0}(q^2)$ and $f_{T}(q^2)$. 
In this study, the form factors are parametrized following the BSZ parametrization~\cite{Bharucha_2016}, which is truncated at the second order. The eight resulting hadronic parameters are obtained from a joint theoretical prior probability density function (PDF). This PDF is comprised of the 2021 lattice world average based on results by the Fermilab/MILC and HPQCD collaborations~\cite{FlavourLatticeAveragingGroupFLAG:2021npn,Parrott:2022rgu}. Theoretical predictions are obtained from the \texttt{EOS} software~\cite{EOSAuthors:2021xpv,EOS:v1.0.16}. 

\begin{figure}
    \centering
    \includegraphics[width=\linewidth]{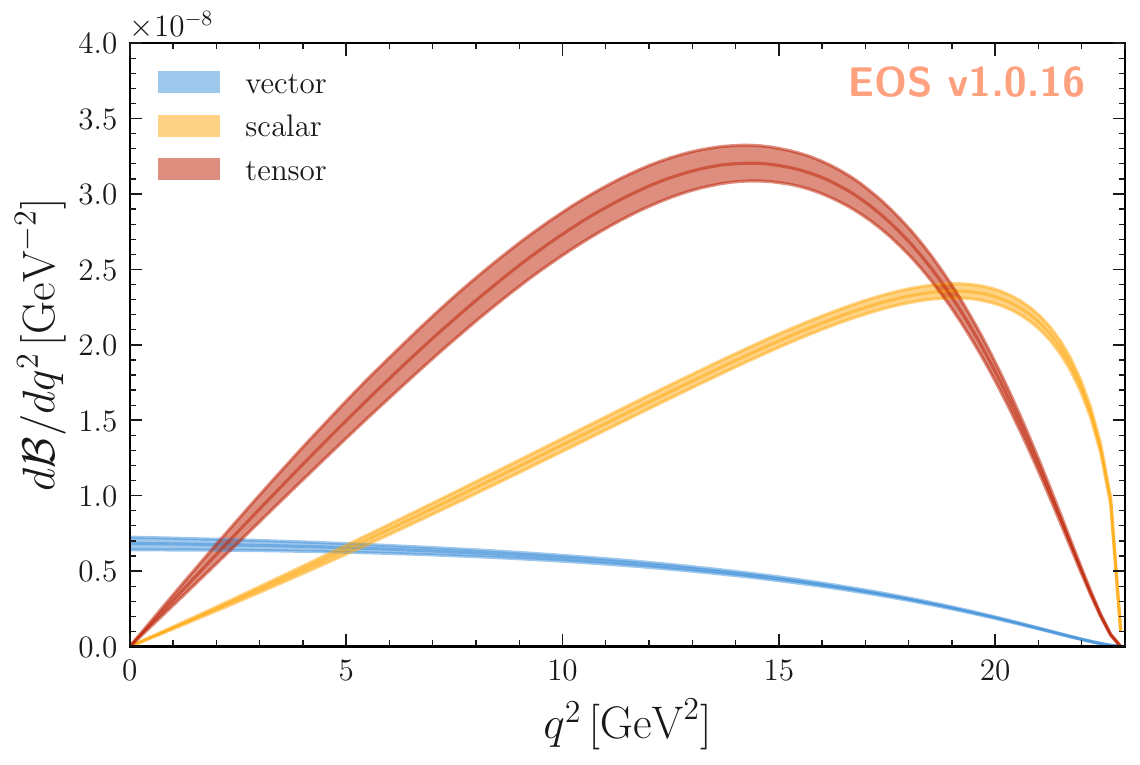}
    \caption{The \BKnn differential branching fraction prediction from \cref{eq:wet-width}. Individual contributions are shown here with the combinations of vector, scalar and tensor Wilson coefficients set to unity, respectively. The uncertainties shown (bands) stem from the hadronic parameters.}
    \label{fig:wet-theory}
\end{figure}

The predicted kinematic distributions of the respective vector, scalar and tensor operators are shown in \cref{fig:wet-theory}.

\section{Results}

\subsection{Inference of Wilson coefficients}

To obtain a marginal posterior for the WET Wilson coefficients, we introduce 11 additional parameters to the \BKnn statistical model. These include three unconstrained parameters of interest,
\begin{equation}
    \boldsymbol{\eta} = [C_{\mathrm{VL}}+C_{\mathrm{VR}}, ~C_{\mathrm{SL}}+C_{\mathrm{SR}}, ~C_{\mathrm{TL}}]\, ,
\end{equation}
along with eight nuisance parameters that parameterize the hadronic form factors. The latter set comprises 8 correlated parameters, which are decorrelated using the eigendecomposition of their covariance matrix (see App.~B of Ref.~\cite{Gartner:2024muk}). The three nuisance parameters for the hadronic parameters entering the SM prediction, which were already present in the statistical model, are removed to avoid double counting.

\begin{figure}
    \centering
    \includegraphics[width=0.5\textwidth]{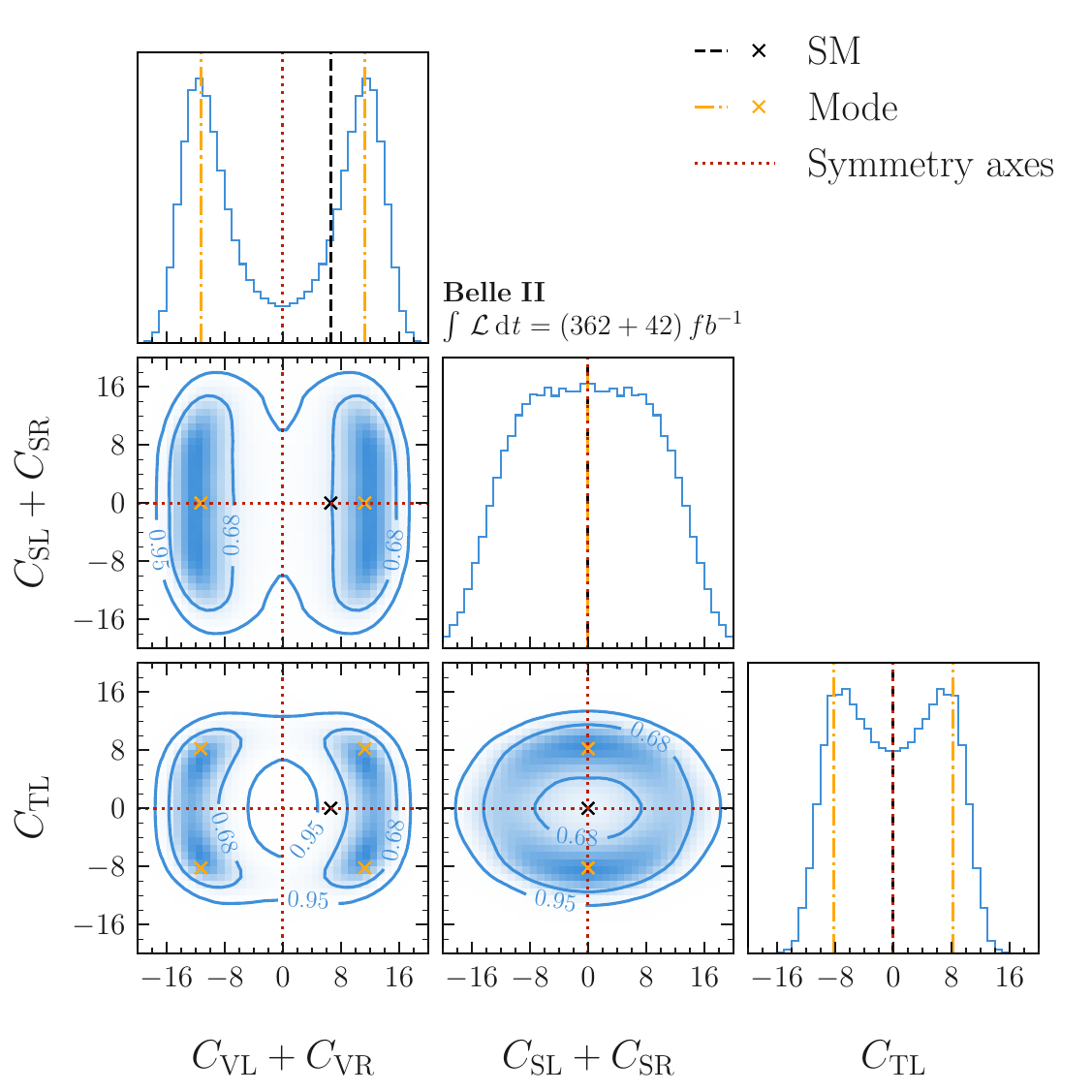}
    \caption{The marginalized posterior for the Wilson coefficients in \cref{eq:wet-width}. We adopt the convention that $C_{\mathrm{VL}}+C_{\mathrm{VR}}$, $C_{\mathrm{SL}}+C_{\mathrm{SR}}$ and $C_{\mathrm{TL}}$ are real valued.
    Diagonal and off-diagonal panels show the 1-dimensional and 2-dimensional sample density PDFs on a linear scale, respectively. The overall scale is omitted, as all relevant information is contained in the shape of the distribution. The contours indicate $68\%$ and $95\%$ credible intervals. The dashed black lines and cross mark the SM point; the dash-dotted yellow lines and cross indicate the posterior mode; dotted red lines mark the symmetry axes used for sample symmetrization.
    }
    \label{fig:wet-posterior}
\end{figure}

We exploit the symmetry of \cref{eq:wet-width} and sample only in the octant of the parameter space where all Wilson coefficients are positive, and symmetrize the samples afterward. We choose uniform priors for all Wilson coefficients in the range $[0,~20]$. Uniform priors are justified by neither wanting to assign preference to any part of the parameter space, nor anticipating inference based on a non-linear transformation of the Wilson coefficients. Ranges are chosen to cover the full posterior.
The marginal posterior is shown in \cref{fig:wet-posterior}. 
There is a clear deviation from the SM in the vector sector, as expected from the result of Ref.~\cite{Belle-II:2023esi}.
Further, we find that the posterior distribution peaks around a non-zero value for the tensor contribution. This indicates that a pure SM signal template does not provide the best description of the data (see \cref{sec:post fit yields}).
From the 1-dimensional marginal posterior distributions, we can calculate the highest density credible intervals (HDI)\footnote{The smallest possible credible interval at a given probability level.} at 68\% and 95\% probability on the absolute values of the Wilson coefficients. The posterior mode and the credible intervals are shown in \cref{tab:wet results}.
\begin{table}[ht]
    \renewcommand{\arraystretch}{1.5}
    \caption{The mode of the posterior, and HDI at 68\% and 95\% for the (sums of the) WET Wilson coefficients in \cref{eq:wet-width}, derived from the posterior in \cref{fig:wet-posterior}.}
    \centering
    \begin{tabular*}{\linewidth}{@{\extracolsep{\fill}}lccc@{}}
        \toprule \midrule
        \textbf{Parameters} & \textbf{Mode} & \textbf{68\% HDI} & \textbf{95\% HDI} \\
        \midrule
         $
         \begin{aligned}
            &|C_\mathrm{VL}+C_\mathrm{VR}|\\
            &|C_\mathrm{SL}+C_\mathrm{SR}|\\
            &|C_\mathrm{TL}|
        \end{aligned}
        $ & $
        \begin{aligned}
            11.3&\\
            0.0& \\
            8.2&
        \end{aligned}
        $ & $
         \begin{aligned}
            [7.8, ~14.6]&\\
            [0.0, ~9.6]& \\
            [2.3, ~9.6]&
        \end{aligned}
        $ & $
        \begin{aligned}
            &[1.9, ~16.2]\\
            &[0.0, ~15.4] \\
            &[0.0, ~11.2]
        \end{aligned}
        $ \\
        \midrule \bottomrule
    \end{tabular*}
    \label{tab:wet results}
\end{table}

To provide a baseline for comparison, we assess the effect of neglecting kinematic shape information by computing credible intervals using a simplified reinterpretation approach. 
In this approach, the SM branching ratio is scaled by an overall factor, discarding all differential distribution information. This is implemented by performing inference with a likelihood constructed from joint number densities defined over a single bin in the kinematic range, reducing \cref{eq:reweight_discrete} to $\nu_{1,x} = \nu_{0,x} w$.
Within this simplified framework, the resulting 95\% credible intervals are
${|C_\mathrm{VL}+C_\mathrm{VR}| < 14.4}$, ${|C_\mathrm{SL}+C_\mathrm{SR}|< 8.5}$ and ${|C_\mathrm{TL}|< 7.1}$.
These values demonstrate a bias from neglecting kinematic shape differences, as evidenced by the discrepancies compared to the results in \cref{tab:wet results}. The significant variations highlight the importance of reinterpretation approaches that account for these kinematic shape differences.

A prior sensitivity analysis was performed to assess the dependence of the presented results on the chosen priors (see \cref{sec:sensitivity}). Two alternative prior choices were considered. The most stable parameter was found to be $ |C_\mathrm{VL} + C_\mathrm{VR}| $, while the largest variations occurred for $ |C_\mathrm{SL} + C_\mathrm{SR}| $, due to the low sensitivity of the analysis to this parameter.

\subsection{Model comparison}

We compare the goodness of fit and relative performance of the WET and the unconstrained \BKnn SM, where the signal strength parameter $\mu_{\rm SM}$ is treated as a free parameter.
To assess relative performance, these models are evaluated against the background-only (BKG) hypothesis, which assumes no \BKnn contribution, and the constrained \BKnn SM, where $\mu_{\rm SM}$ is fixed to the SM prediction. The constrained model includes a 5\% normalization uncertainty, which accounts for a $4.4\%$ uncertainty in the CKM matrix elements $|V_{ts}^*V^{}_{tb}|^2$ and a $2.3\%$ uncertainty in $|C_{\rm VL}^{\rm SM}|^2$ \cite{Parrott:2022zte}.

A detailed summary of the models and their defining characteristics is provided in \cref{tab:models}.

\begin{table}[ht]
    \renewcommand{\arraystretch}{1.5}
    \caption{A summary of the models under consideration, their corresponding theoretical predictions, references for the hadronic parameters, and the number of hadronic parameters included in each model is provided.}
    \centering
    \begin{tabular*}{\linewidth}{@{\extracolsep{\fill}}lllc@{}}
        \toprule \midrule
        \textbf{Model} & \textbf{Prediction} & \multicolumn{2}{c}{\textbf{Hadronic Params.}} \\
        & & \textbf{Ref.} & \textbf{Nr.} \\
        \midrule
        WET & \cref{eq:wet-width} & \cite{FlavourLatticeAveragingGroupFLAG:2021npn,Parrott:2022rgu} &  8\\ \addlinespace \addlinespace
        SM unconstrained & \makecell[l]{\cref{eq:wet-width}, $C_{\rm VL} \geq 0$ \\ $C_{\gamma \delta}=0$ otherwise} & \cite{Parrott:2022rgu} & 3\\ \addlinespace \addlinespace
        SM constrained & \makecell[l]{\cref{eq:wet-width}, $C_{\rm VL}=6.6$,\\ $C_{\gamma \delta}=0$ otherwise, \\ 5\% norm. unc.} & \cite{Parrott:2022rgu} & 3\\ \addlinespace \addlinespace
        BKG & no signal & -- & --\\
        \midrule \bottomrule
    \end{tabular*}
    \label{tab:models}
\end{table}

A local goodness-of-fit (gof) $P$-value is calculated from 
\begin{equation}
        P_{\rm gof} = \int_{t_{\rm obs}}^\infty dt ~ p(t), \quad
        t = -2 \ln \frac{p(\boldsymbol{n}, \boldsymbol{a} \mid \hat{\boldsymbol{\eta}}, \hat{ \boldsymbol{\chi}})}{p_{\rm sat}(\boldsymbol{n}, \boldsymbol{a} \mid \bar{\boldsymbol{\chi}})}\,,
\end{equation}
where $\hat{\boldsymbol{\eta}}, \hat{ \boldsymbol{\chi}}$ is the best-fit point, and $p_{\rm sat}(\boldsymbol{n}, \boldsymbol{a} \mid \bar{\boldsymbol{\chi}})$ is the saturated likelihood. In the saturated likelihood, the expected event rates of the \texttt{Histfactory} likelihood~\cite{histfactory} are set to the observed data, with all constraint terms maximized at $\bar \chi$.
The PDF of the test statistic $p(t)$ is obtained from fits to toy data,\footnote{Note that for the WET, $p(t)$ does not follow the asymptotic chi-square distribution, as the Wilson coefficients can only positively contribute to the rate in \cref{eq:wet-width} (for a discussion see Ref.~\cite{Bernlochner:2022oiw}).} sampled from
$
    p(\boldsymbol{n}, \boldsymbol{a} \mid \hat{\boldsymbol{\eta}}, \hat{ \boldsymbol{\chi}}).
$
The goodness of fit $P$-values are reported in \cref{tab:bayes factors}, indicating good fits for both models.


A global model comparison is performed using the Bayes factor, defined as the ratio of the marginal likelihoods between two competing models. The computed Bayes factors are presented in \cref{tab:bayes factors}. Both models exceed Jeffreys' condition~\cite{jeffreys1961theory} for a \textit{very strong} model preference ($\log_{10} B > 1.5$) over the background-only hypothesis. Furthermore, both models surpass the threshold for a \textit{substantial} model preference ($\log_{10} B > 0.5$) over the constrained \BKnn SM.

\begin{table}[ht]
    \renewcommand{\arraystretch}{1.5}
    \caption{The Bayes factors of each model over the background (constrained \BKnn SM) hypothesis, $B_{\rm BKG}$ ($B_{\rm SM}^{\rm constr.}$), and the goodness of fit $P$-values.}
    \centering
    \begin{tabular*}{\linewidth}{@{\extracolsep{\fill}}lccc@{}}
        \toprule \midrule
        \textbf{Model} & $\boldsymbol{\log_{10} B_{\rm BKG}}$ & $\boldsymbol{\log_{10} B_{\rm SM}^{\rm constr.}}$ & $\boldsymbol{P_{\rm gof}}$\\
        \midrule
    WET & 1.8 & 0.7 & 0.6\\
    SM unconstrained  & 2.0 & 0.9 & 0.6\\
        \midrule \bottomrule
    \end{tabular*}
    \label{tab:bayes factors}
\end{table}

To provide a complementary frequentist perspective and support the conclusions derived from the Bayesian model comparison in \cref{tab:bayes factors}, a hypothesis test was conducted using the $P$-value
\begin{equation}
        P = \int_{t_{\rm obs}}^\infty dt ~ p(t), \quad
        t = -2 \ln \frac{p(\boldsymbol{n}, \boldsymbol{a} \mid \boldsymbol{\eta} = \boldsymbol{0}, \hat{\hat{ \boldsymbol{\chi}}})}{p(\boldsymbol{n}, \boldsymbol{a} \mid \hat{\boldsymbol{\eta}}, \hat{ \boldsymbol{\chi}})}\,.
\end{equation}
While the Bayes factor compares models globally across parameter space, the $P$-value provides a local test of statistical significance at the best-fit point.
This test compares the model at $\hat{\boldsymbol{\eta}}, \hat{ \boldsymbol{\chi}}$ against the background-only hypothesis, where $\boldsymbol{\eta} = \boldsymbol{0}$ and the nuisance parameters are optimized to $\hat{\hat{ \boldsymbol{\chi}}}$.
The distribution of the test statistic $p(t)$ is obtained from fits to toy data, sampled from $p(\boldsymbol{n}, \boldsymbol{a} \mid \boldsymbol{\eta} = \boldsymbol{0}, \hat{\hat{ \boldsymbol{\chi}}})$.
For the WET we obtain a $P$-value of $P=4.6\cdot 10^{-4}$, corresponding to a significance of $Z=3.3$, over the background-only hypothesis. This is smaller than the $Z=3.5$ obtained in Ref.~\cite{Belle-II:2023esi} because the present analysis uses a likelihood ratio test statistic rather than an upper limit test statistic.

\section{Conclusion}

In conclusion, this paper describes a robust reinterpretation of the \BKnn result~\cite{Belle-II:2023esi} within the WET, using the model-agnostic likelihood approach. From the analysis likelihood, in conjunction with the joint number densities, we construct Bayesian posterior models and derive marginal posteriors as well as credible intervals on theoretical model parameters. 

This work presents the first reported credible intervals on the $b\to s$ WET Wilson coefficients from a thorough reinterpretation of Belle~II data (see \cref{tab:wet results}). The posterior mode for the magnitudes of the Wilson coefficient combinations is found at ${(|C_\mathrm{VL}+C_\mathrm{VR}|,\, |C_\mathrm{SL}+C_\mathrm{SR}|,\, |C_\mathrm{TL}|) = (11.3,\, 0.0,\, 8.2)}$, with corresponding 95\% credible intervals of $[1.9,\, 16.2]$, $[0.0,\, 15.4]$, and $[0.0,\, 11.2]$, respectively. The results indicate that an enhancement in the vector sector is required to obtain good compatibility with the data. However, the best overall compatibility is achieved with an additional, sizable tensor contribution.

A central goal of this work is to enable broad reinterpretation of the \BKnn measurement by publishing the model-agnostic likelihood. This consists of the full likelihood from the \BKnn result~\cite{Belle-II:2023esi} and the joint number densities used in this reinterpretation~\cite{hepdata.146803, hepdata.166082} (for details see \cref{sec:BelleIIdata}). This allows the broader scientific community to test alternative theoretical models in a statistically rigorous way. 
Importantly, this publication sets a template for future Belle~II measurements that are suitable for reinterpretation and reflects the collaboration's commitment to publish model-agnostic likelihoods as a means to maximize the scientific impact and reusability of its results.

This work, based on data collected using the Belle II detector, which was built and commissioned prior to March 2019,
was supported by
Higher Education and Science Committee of the Republic of Armenia Grant No.~23LCG-1C011;
Australian Research Council and Research Grants
No.~DP200101792, 
No.~DP210101900, 
No.~DP210102831, 
No.~DE220100462, 
No.~LE210100098, 
and
No.~LE230100085; 
Austrian Federal Ministry of Education, Science and Research,
Austrian Science Fund (FWF) Grants
DOI:~10.55776/P34529,
DOI:~10.55776/J4731,
DOI:~10.55776/J4625,
DOI:~10.55776/M3153,
and
DOI:~10.55776/PAT1836324,
and
Horizon 2020 ERC Starting Grant No.~947006 ``InterLeptons'';
Natural Sciences and Engineering Research Council of Canada, Compute Canada and CANARIE;
National Key R\&D Program of China under Contract No.~2024YFA1610503,
and
No.~2024YFA1610504
National Natural Science Foundation of China and Research Grants
No.~11575017,
No.~11761141009,
No.~11705209,
No.~11975076,
No.~12135005,
No.~12150004,
No.~12161141008,
No.~12475093,
and
No.~12175041,
and Shandong Provincial Natural Science Foundation Project~ZR2022JQ02;
the Czech Science Foundation Grant No. 22-18469S,  Regional funds of EU/MEYS: OPJAK
FORTE CZ.02.01.01/00/22\_008/0004632 
and
Charles University Grant Agency project No. 246122;
European Research Council, Seventh Framework PIEF-GA-2013-622527,
Horizon 2020 ERC-Advanced Grants No.~267104 and No.~884719,
Horizon 2020 ERC-Consolidator Grant No.~819127,
Horizon 2020 Marie Sklodowska-Curie Grant Agreement No.~700525 ``NIOBE''
and
No.~101026516,
and
Horizon 2020 Marie Sklodowska-Curie RISE project JENNIFER2 Grant Agreement No.~822070 (European grants);
L'Institut National de Physique Nucl\'{e}aire et de Physique des Particules (IN2P3) du CNRS
and
L'Agence Nationale de la Recherche (ANR) under Grant No.~ANR-21-CE31-0009 (France);
BMFTR, DFG, HGF, MPG, and AvH Foundation (Germany);
Department of Atomic Energy under Project Identification No.~RTI 4002,
Department of Science and Technology,
and
UPES SEED funding programs
No.~UPES/R\&D-SEED-INFRA/17052023/01 and
No.~UPES/R\&D-SOE/20062022/06 (India);
Israel Science Foundation Grant No.~2476/17,
U.S.-Israel Binational Science Foundation Grant No.~2016113, and
Israel Ministry of Science Grant No.~3-16543;
Istituto Nazionale di Fisica Nucleare and the Research Grants BELLE2,
and
the ICSC – Centro Nazionale di Ricerca in High Performance Computing, Big Data and Quantum Computing, funded by European Union – NextGenerationEU;
Japan Society for the Promotion of Science, Grant-in-Aid for Scientific Research Grants
No.~16H03968,
No.~16H03993,
No.~16H06492,
No.~16K05323,
No.~17H01133,
No.~17H05405,
No.~18K03621,
No.~18H03710,
No.~18H05226,
No.~19H00682, 
No.~20H05850,
No.~20H05858,
No.~22H00144,
No.~22K14056,
No.~22K21347,
No.~23H05433,
No.~26220706,
and
No.~26400255,
and
the Ministry of Education, Culture, Sports, Science, and Technology (MEXT) of Japan;  
National Research Foundation (NRF) of Korea Grants
No.~2021R1-F1A-1064008, 
No.~2022R1-A2C-1003993,
No.~2022R1-A2C-1092335,
No.~RS-2016-NR017151,
No.~RS-2018-NR031074,
No.~RS-2021-NR060129,
No.~RS-2023-00208693,
No.~RS-2024-00354342
and
No.~RS-2025-02219521,
Radiation Science Research Institute,
Foreign Large-Size Research Facility Application Supporting project,
the Global Science Experimental Data Hub Center, the Korea Institute of Science and
Technology Information (K25L2M2C3 ) 
and
KREONET/GLORIAD;
Universiti Malaya RU grant, Akademi Sains Malaysia, and Ministry of Education Malaysia;
Frontiers of Science Program Contracts
No.~FOINS-296,
No.~CB-221329,
No.~CB-236394,
No.~CB-254409,
and
No.~CB-180023, and SEP-CINVESTAV Research Grant No.~237 (Mexico);
the Polish Ministry of Science and Higher Education and the National Science Center;
the Ministry of Science and Higher Education of the Russian Federation
and
the HSE University Basic Research Program, Moscow;
University of Tabuk Research Grants
No.~S-0256-1438 and No.~S-0280-1439 (Saudi Arabia), and
Researchers Supporting Project number (RSPD2025R873), King Saud University, Riyadh,
Saudi Arabia;
Slovenian Research Agency and Research Grants
No.~J1-50010
and
No.~P1-0135;
Ikerbasque, Basque Foundation for Science,
State Agency for Research of the Spanish Ministry of Science and Innovation through Grant No. PID2022-136510NB-C33, Spain,
Agencia Estatal de Investigacion, Spain
Grant No.~RYC2020-029875-I
and
Generalitat Valenciana, Spain
Grant No.~CIDEGENT/2018/020;
The Knut and Alice Wallenberg Foundation (Sweden), Contracts No.~2021.0174 and No.~2021.0299;
National Science and Technology Council,
and
Ministry of Education (Taiwan);
Thailand Center of Excellence in Physics;
TUBITAK ULAKBIM (Turkey);
National Research Foundation of Ukraine, Project No.~2020.02/0257,
and
Ministry of Education and Science of Ukraine;
the U.S. National Science Foundation and Research Grants
No.~PHY-1913789 
and
No.~PHY-2111604, 
and the U.S. Department of Energy and Research Awards
No.~DE-AC06-76RLO1830, 
No.~DE-SC0007983, 
No.~DE-SC0009824, 
No.~DE-SC0009973, 
No.~DE-SC0010007, 
No.~DE-SC0010073, 
No.~DE-SC0010118, 
No.~DE-SC0010504, 
No.~DE-SC0011784, 
No.~DE-SC0012704, 
No.~DE-SC0019230, 
No.~DE-SC0021274, 
No.~DE-SC0021616, 
No.~DE-SC0022350, 
No.~DE-SC0023470; 
and
the Vietnam Academy of Science and Technology (VAST) under Grants
No.~NVCC.05.12/22-23
and
No.~DL0000.02/24-25.

These acknowledgements are not to be interpreted as an endorsement of any statement made
by any of our institutes, funding agencies, governments, or their representatives.

We thank the SuperKEKB team for delivering high-luminosity collisions;
the KEK cryogenics group for the efficient operation of the detector solenoid magnet and IBBelle on site;
the KEK Computer Research Center for on-site computing support; the NII for SINET6 network support;
and the raw-data centers hosted by BNL, DESY, GridKa, IN2P3, INFN, 
and the University of Victoria.

\appendix

\section{Predicted yields at the posterior mode}
\label{sec:post fit yields}

\begin{figure}
    \centering
    \includegraphics[width=\linewidth]{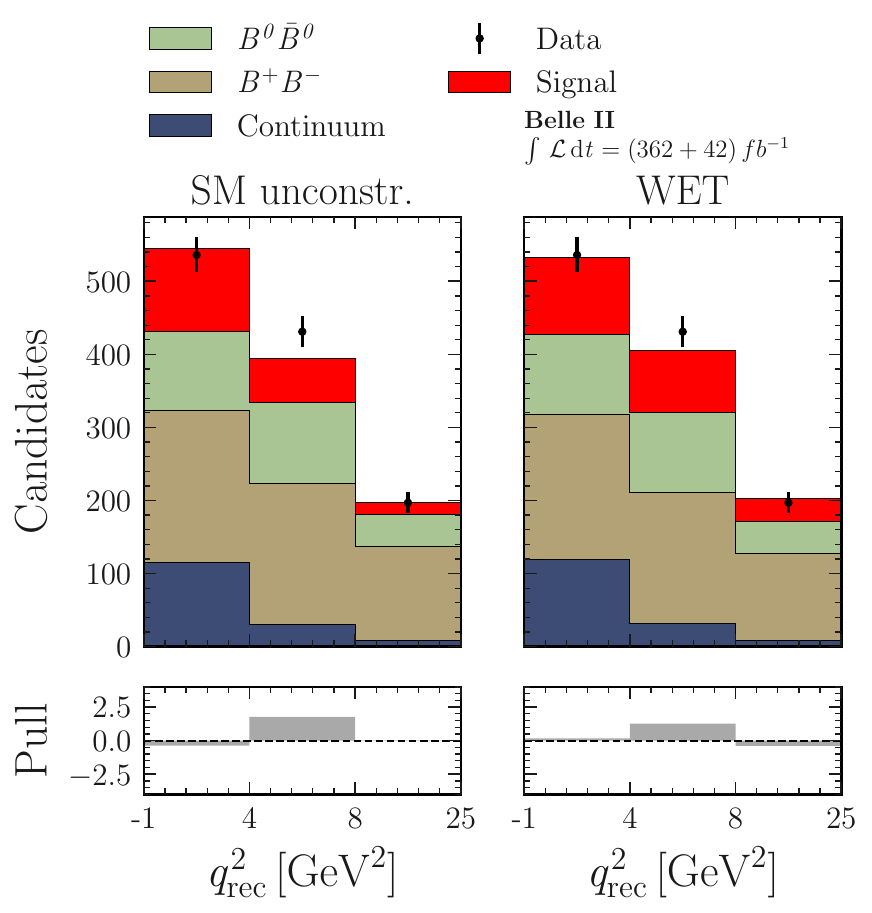}
    \caption{Observed and predicted best-fit yields in the highest-sensitivity bins of the analysis. These correspond to the $\eta(\BDT2)>0.98$ region of the ITA. The signal is shown for the unconstrained \BKnn SM (\textit{left}) and the WET (\textit{right}) predictions.
    The predicted background yields are shown individually for the neutral and charged \B-meson decays, and the summed five continuum categories. 
    Pulls are shown in the lower panels.
    }
    \label{fig:post-fit}
\end{figure}

Direct comparison of the observed data yields to the predicted yields at the posterior mode parameter point (the best-fit point to data; see \cref{tab:wet results}) for the unconstrained \BKnn SM and the WET is shown in \cref{fig:post-fit}, for the highest-sensitivity bins of the analysis. 
The WET model provides a better fit to the data than the unconstrained SM prediction, as indicated by the smaller pull values.

\section{Prior sensitivity study}
\label{sec:sensitivity}
To investigate the sensitivity of the results in \cref{tab:wet results} to the choice of priors, we derived the posterior mode and credible intervals for alternative sets of priors.

First, we select truncated-normal priors centered on the SM expectation (the only non-zero Wilson coefficient being $C_{\mathrm{VL}}^{\mathrm{SM}}=6.6$), which disfavor deviations from the SM expectation,
\begin{equation}
p\left( \eta_i \right) = 
\begin{cases}
    \mathcal{N}(\eta_i | \mu=C_i^{\mathrm{SM}}, \sigma=20) &~ \eta_i \geq 0 \\
    0 &~ \eta_i < 0
\end{cases}.
\label{eq:wet priors normal}
\end{equation}
Here, $\eta_i \in [|C_\mathrm{VL}+C_\mathrm{VR}|, |C_\mathrm{SL}+C_\mathrm{SR}|, |C_\mathrm{TL}|]$ and $C_i^{\mathrm{SM}}$ correspond to the respective SM point $C_i^{\mathrm{SM}} \in [6.6, 0.0, 0.0]$. 

Second, we select uniform priors in the squared Wilson coefficients, as these enter \cref{eq:wet-width}, which subsequently translate to linear priors for the Wilson coefficients,
\begin{equation}
    p\left( \eta_i \right) \propto  \begin{cases}
        \eta_i &~ \eta_i \leq 30\\
        0 &~ \eta_i > 30
        \end{cases}.
    \label{eq:wet priors triangular}
\end{equation}
These priors favor larger values for the Wilson coefficients.

The resulting credible intervals for both cases are shown in \cref{tab:wet results sensitivity}. The vector Wilson coefficient posterior mode and credible intervals are found to be the most robust to prior choices. The largest changes are found for the scalar Wilson coefficients, to which the analysis is the least sensitive, due to low efficiency at high $q^2$. This is also expected from the posterior distribution in \cref{fig:wet-posterior}.

\begin{table}[ht]
    \renewcommand{\arraystretch}{1.5}
    \caption{The posterior modes, and HDIs at 68\% and 95\% for the (sums of the) WET Wilson coefficients in \cref{eq:wet-width}, for alternative prior choices (cf. \cref{tab:wet results}). }
    \centering
    \begin{tabular*}{\linewidth}{@{\extracolsep{\fill}}llccc@{}}
        \toprule \midrule
        \textbf{Priors} & \textbf{Parameters} & \textbf{Mode} & \textbf{68\% HDI} & \textbf{95\% HDI} \\
        \midrule
        \cref{eq:wet priors normal} &
         $
         \begin{aligned}
            &|C_\mathrm{VL}+C_\mathrm{VR}|\\
            &|C_\mathrm{SL}+C_\mathrm{SR}|\\
            &|C_\mathrm{TL}|
        \end{aligned}
        $ & $
        \begin{aligned}
              11.4&\\
              0.0& \\
              7.7&
        \end{aligned}
        $ & $
         \begin{aligned}
            [8.0, 14.6]&\\
            [0.0, 9.2]& \\
            [1.5, 8.8]&
        \end{aligned}
        $ & $
        \begin{aligned}
              &[2.2, 16.4]\\
              &[0.0, 14.7] \\
              &[0.0, 11.0]
        \end{aligned}
        $
        \\
        \midrule
        \cref{eq:wet priors triangular} &
         $
         \begin{aligned}
            &|C_\mathrm{VL}+C_\mathrm{VR}|\\
            &|C_\mathrm{SL}+C_\mathrm{SR}|\\
            &|C_\mathrm{TL}|
        \end{aligned}
        $ & $
        \begin{aligned}
              11.6&\\
              8.9&\\
              7.2&
        \end{aligned}
        $ & $
         \begin{aligned}
            [8.2, 14.0]&\\
            [4.6, 12.6]& \\
            [3.9, 9.6]&
        \end{aligned}
        $ & $
        \begin{aligned}
              [4.2, 16.0]&\\
              [1.3, 15.6]& \\
              [1.4, 11.7]&
        \end{aligned}
        $\\
        \midrule \bottomrule
    \end{tabular*}
    \label{tab:wet results sensitivity}
\end{table}

\section{HEPData inventory}
\label{sec:BelleIIdata}

To enable reinterpretation under any NP model with the model-agnostic likelihood~\cite{Gartner:2024muk}, the necessary information from Belle~II will be published on \href{https://www.hepdata.net/}{HEPData}~\cite{hepdata.146803,hepdata.166082}. The release will include the following components:

\begin{itemize}
\item The SM \BKnn differential branching fraction as a function of $q^{2}$;
\item Signal selection efficiency as a function of $q^{2}$;
\item Binned joint number densities:
\begin{itemize}
    \item ITA: x-axis: $q^{2}$, y-axis: $q^{2}_{\rm{rec}} \times \eta(\rm{BDT}_{2})$ (flattened), z-axis: events (weighted);
    \item HTA: x-axis: $q^{2}$, y-axis: $\eta(\rm{BDTh})$ (flattened), z-axis: events (weighted);
\end{itemize}  
\item \texttt{pyhf} combined likelihood in \texttt{json} format:
\begin{itemize}
\item Containing templates for signal and background after all selections, binned in ${q^{2}_{\rm{rec}} \times \eta(\rm{BDT}_{2})}$ (ITA) and $\eta(\rm{BDTh})$ (HTA);
\end{itemize}  
\item The code to reproduce the WET reinterpretation results obtained in this analysis.
\end{itemize}
\bibliographystyle{apsrev4-2}
\bibliography{references}

\end{document}